\documentclass{PoS}

\title{Continuum extrapolation of the critical endpoint in 4-flavor QCD with Wilson-Clover fermions}

\ShortTitle{Continuum extrapolation of the critical endpoint in 4-flavor QCD with Wilson-Clover fermions}

\author{\speaker{Hiroshi Ohno}\\
        Center for Computational Sciences, Univesity of Tsukuba, Tsukuba, Ibaraki 305-8577, Japan\\
        E-mail: \email{hohno@ccs.tsukuba.ac.jp}}

\author{Yoshinobu Kuramashi\\
        Center for Computational Sciences, Univesity of Tsukuba, Tsukuba, Ibaraki 305-8577, Japan\\
        E-mail: \email{kuramasi@het.ph.tsukuba.ac.jp}}

\author{Yoshifumi Nakamura\\
        RIKEN Center for Computational Science, Kobe, Hyogo 650-0047, Japan\\
        E-mail: \email{nakamura@riken.jp}}

\author{Shinji Takeda\\
        Institute of Physics, Kanazawa University, Kanazawa 920-1192, Japan\\
        E-mail: \email{takeda@hep.s.kanazawa-u.ac.jp}}

\abstract{We report our study on the critical endpoint of the finite temperature phase transition in 4-flavor QCD
with Wilson-Clover fermions. Using the kurtosis intersection method, we determined the critical endpoint on lattices with
$N_t$ = 4, 6 and 8. Our continuum extrapolated results show that the pseudo-scalar meson mass at the critical endpoint,
$m_\mathrm{PS,E}$, for 4-flavor is clearly larger than that for 3-flavor. We also compared our results to those with staggered
fermions and found that $m_\mathrm{PS,E}/T_E$ for 4-flavor with Wilson fermions might remain finite even in the continuum limit in contrast
to that with staggered fermions, where $m_\mathrm{PS,E}/T_E$ is very close to zero, which suggests that the difference between Wilson and
staggered fermions is at least not due to the rooting.}

\FullConference{The 36th Annual International Symposium on Lattice Field Theory - LATTICE2018\\
		22-28 July, 2018\\
		Michigan State University, East Lansing, Michigan, USA.}

\begin{document}

\section{Introduction}
Understanding the nature of finite temperature phase transition is an important subject in Quantum chromodynamics (QCD).
It is known that the order of the phase transition depends on the quark mass $m_q$ as well as the number of quark flavors $N_f$ \cite{Pisarski:1983ms}.
Especially in the case of three degenerate massless flavors, the phase transition is expected to be of first order. It is also expected that
for a certain large $m_q$ there is no longer any phase transition but an analytic crossover. Therefore, as increasing $m_q$ from zero,
one should find a second order phase transition point or a so-called critical endpoint which separates regions of the first order phase transition
and the crossover.

However, to pin down the critical endpoint of $N_f = 3$ QCD is a longstanding issue in the lattice QCD community. Previous studies with staggered-type
quarks \cite{Karsch:2001nf, Karsch:2003va, deForcrand:2007rq, Varnhorst:2015lea, Bazavov:2017xul} have shown that the critical pion mass,
a pion mass at the critical endpoint, gets smaller as using more improved fermion actions and smaller lattice spacings, which suggests that the critical
pion mass in the continuum limit is quite small or consistent with zero. A similar tendency has been also reported with Wilson-type quarks
\cite{Iwasaki:1996zt, Jin:2014hea, Jin:2017jjp} but the critical pion mass is always much larger than that for the staggered ones at any lattice spacing and
even in the continuum limit. Therefore, following two questions are naturally raised: one is whether the critical pion mass can be really so small as given with the staggered-type
quarks and the other is why the critical pion mass for the two different types of quarks are so different.

To answer these questions, in this study, we consider $N_f = 4$ QCD as a good analogue of $N_f = 3$ because of following three reasons.
First, a first order phase transition is expected in the massless limit also in $N_f = 4$ \cite{Pisarski:1983ms}. Second, the critical pion mass can be larger in $N_f = 4$
than in $N_f = 3$ based on a naive counting of the degrees of freedom, which means that one needs less computing cost for carrying out a proper continuum
extrapolation in $N_f = 4$ than in $N_f = 3$. Finally, staggered-type quarks are naturally defined in $N_f = 4$ in contrast to $N_f = 3$, where the rooted fermion
determinant, which is a non-trivial operation, is needed to represent three degenerate flavors. Therefore, no concern of the rooting allows better comparison between
staggered- and Wilson-type quarks. Since a study in this context with the naive staggered quarks has already been reported in \cite{deForcrand:2017cgb},
our aim is to find a critical endpoint in $N_f = 4$ with Wilson-type quarks and compare our results with those of \cite{deForcrand:2017cgb}.

\section{Simulation setup}
We employ the Iwasaki gauge and the Wilson-Clover fermion actions, where the clover coefficient is non-perturbatively determined in the same way
as given in a previous study for $N_f =$ 0, 2 and 3 \cite{Aoki:2005et}. Our simulations are performed on lattices with three different lattice spacings
corresponding to temporal extents $N_t$ = 4, 6 and 8 in order to take the continuum limit. We compute the chiral condensate and its cumulants up
to the 4th order, i.e. susceptibility, skewness and kurtosis, by using 10 random sources for each configurations. For each $N_t$ we repeat the calculations
at several hopping parameters $\kappa$ and 2-3 lattice gauge couplings $\beta$ covering the critical endpoint. To improve statistics we also adopt the $\kappa$-reweighting given
in \cite{Kuramashi:2016kpb}. Then, we determine the critical endpoint with the kurtosis intersection method \cite{Karsch:2001nf}, where we check
the finite-size scaling with three different spatial volumes $N_s^3$. BQCD code \cite{Nakamura:2010qh} is used for our simulations.

\section{Results}

\begin{figure}[htbp]
\begin{center}
\includegraphics[width=52mm, angle=-90]{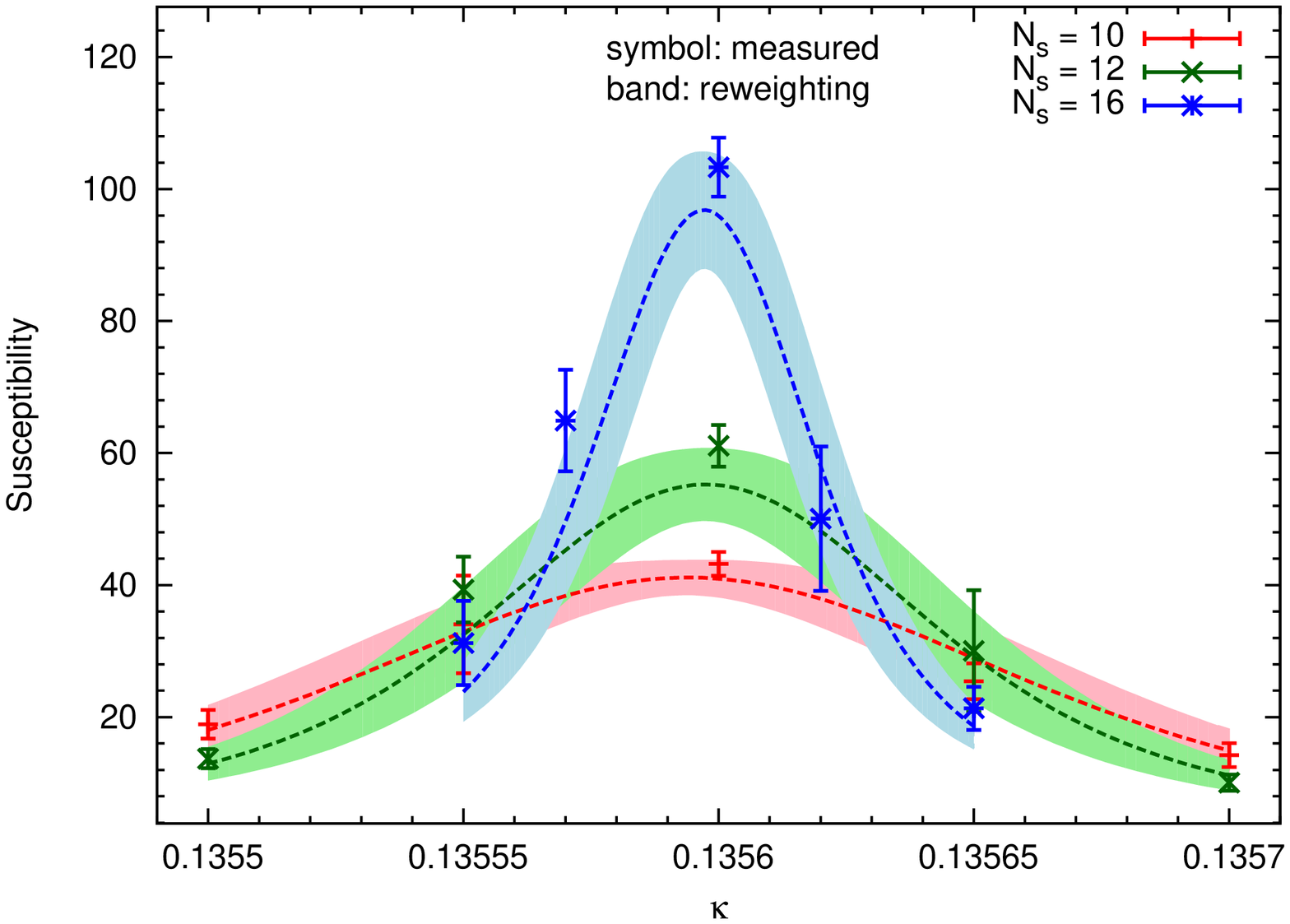}
\includegraphics[width=52mm, angle=-90]{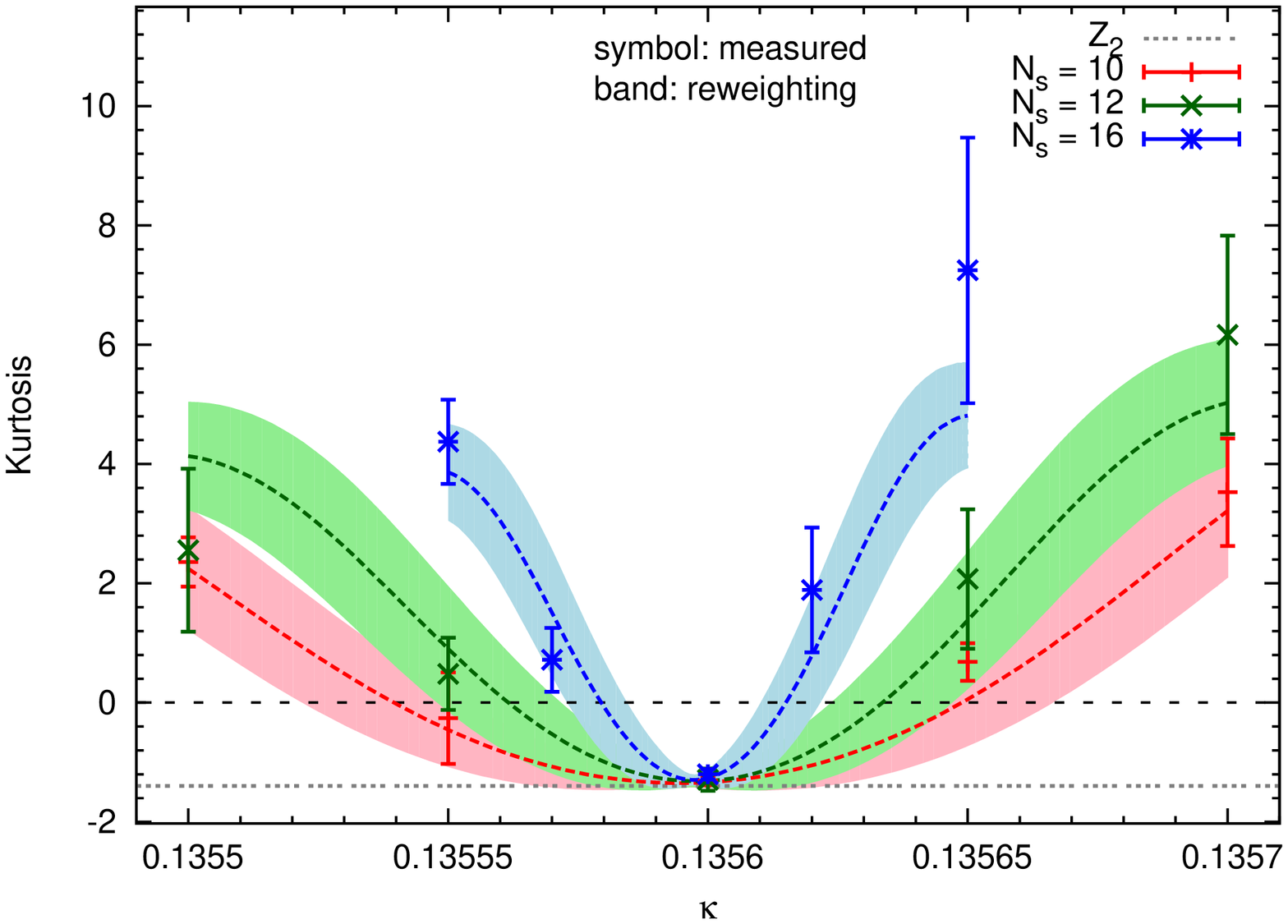}
\caption{An example of the susceptibility (left) and the kurtosis (right) of the chiral condensate. Results at $\beta$ = 1.68 for $N_t$ = 6 are shown.
Difference among colors means different spatial volumes. Data are indicated with symbols. The reweighting results are indicated using dashed curves with
statistical error bands. A kurtosis value at the second order phase transition point belonging to the Z(2) university class is also shown
with a dotted horizontal line in the right panel.}
\label{pbp_cumulants}
\end{center}
\end{figure}
First, an example of the susceptibility and the kurtosis of the chiral condensate is given in Fig. \ref{pbp_cumulants}, where results at $\beta$ = 1.68 for $N_t$ = 6
are shown. It is shown that the reweighting works well and accordingly, clear extrema of the susceptibility and the kurtosis can be seen.

\begin{figure}[htbp]
\begin{center}
\includegraphics[width=65mm, angle=-90]{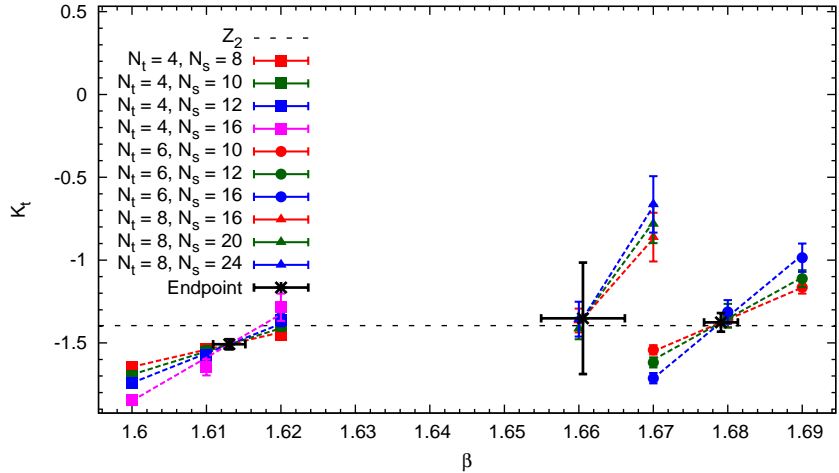}
\caption{Results of the kurtosis intersection method for $N_t = 4$ (squares), 6 (circles) and 8 (triangles). Difference among colors means different spatial volumes.
Critical endpoints for each $N_t$ are indicated with black cross symbols. A kurtosis value at the second order phase transition point belonging to the Z(2) university class
is also shown with a dashed horizontal line.}
\label{pbp_K_intersection}
\end{center}
\end{figure}
Then, Fig. \ref{pbp_K_intersection} shows results for the kurtosis intersection method, where kurtosis values at phase transition/crossover points $K_t$ for each $\beta$
are fitted to an ansatz
\begin{equation}\label{intersection_ansatz}
K_t(\beta) = K_E + cN_s^{1/\nu}(\beta-\beta_E)
\end{equation}
with $\chi^2$/d.o.f. < 0.42, where $\beta_E$ is a $\beta$ value at the critical endpoint. One can see that there are well-defined intersections
indicating the critical endpoints for each $N_t$ and kurtosis values at these points are consistent with one for the Z(2) universality class, -1.396,
except for $N_t = 4$. Note that we do not assume the Z(2) universality for the fits, i.e. both $K_E$ and $\nu$ in Eq. (\ref{intersection_ansatz}) are free parameters.
The critical exponent $\nu$ = 0.7(1) and 0.7(2) for $N_t$ = 4 and 6, respectively, are also close to 0.630 of the Z(2) universality class while more statistics are needed for $N_t$ = 8.  

\begin{figure}[htbp]
\begin{center}
\includegraphics[width=70mm, angle=-90]{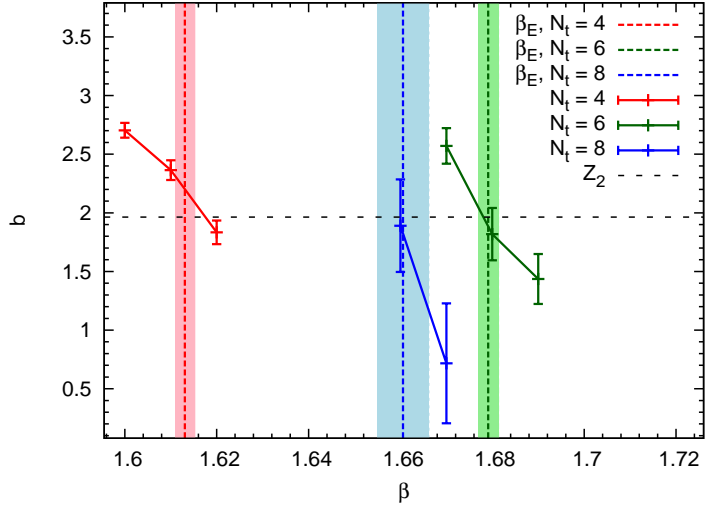}
\caption{$\beta$ dependence of the critical exponent $b$ defined in Eq (\ref{sus_exponent}). Difference among colors means different $N_t$. Vertical dashed lines with bands
indicate $\beta$ values and their statistical errors at the critical endpoints for each $N_t$. $b$ for the Z(2) universality class is also shown with a dashed horizontal line.}
\label{pbp_chi_exponent}
\end{center}
\end{figure}
To further check the universality we also fit the susceptibility peak $\chi_t$ to an ansatz
\begin{equation}\label{sus_exponent}
\chi_t(N_s) = aN_s^{b},
\end{equation}
where we get $\chi^2$/d.o.f. < 0.63. Results of the critical exponent $b$ are shown in Fig. \ref{pbp_chi_exponent}. It can be seen that $b$ at the critical endpoint is consistent with
one for the Z(2) universality class, 1.964, for $N_t$ = 6 and 8 while the inconsistency with the Z(2) universality class is observed again for $N_t = 4$. This might be due to cutoff
effects since we check that our largest volume, 16$^3$, does not influence the location of the critical endpoint within statistical uncertainty, which means that there should not be
any visible finite-size effects.

\begin{figure}[htbp]
\begin{center}
\includegraphics[width=65mm, angle=-90]{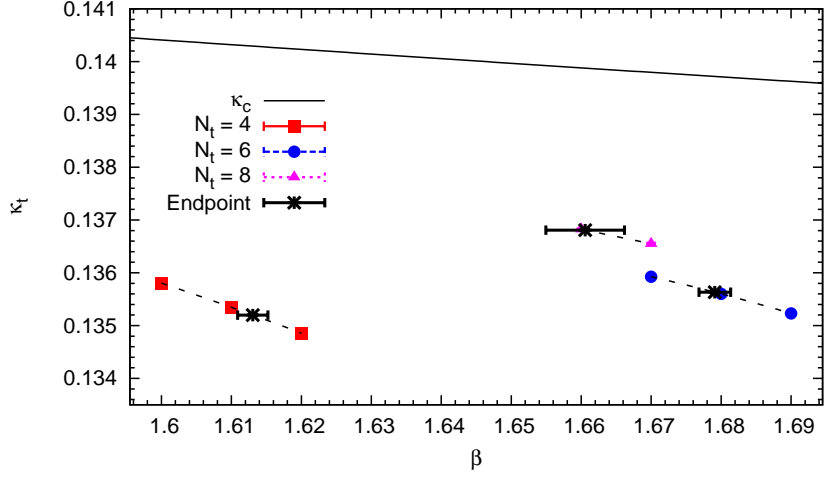}
\caption{$\beta$ dependence of $\kappa_t$. Difference among colors means different $N_t$. Critical endpoints for each $N_t$ are indicated by black cross symbols.
Critical $\kappa$ values $\kappa_c$, where the pion mass becomes zero, are also shown with a solid curve.}
\label{pbp_kappa_t}
\end{center}
\end{figure}
Once $\beta_E$ is given, we determine a $\kappa$ value at the critical endpoint $\kappa_E$ by interpolating $\kappa$ values at the transition/crossover points $\kappa_t$ for each $\beta$
using a quadratic function for $N_t$ = 4 and 6 and a linear function for $N_t$ = 8, where linearly infinite-volume-extrapolated $\kappa$ values are used. The critical endpoints in
the ($\beta$, $\kappa_t$)-plain are shown in Fig. \ref{pbp_kappa_t}, where we find that locations of the critical endpoints have a non-monotonic behavior as increasing $N_t$.

\begin{figure}[htbp]
\begin{center}
\includegraphics[width=150mm]{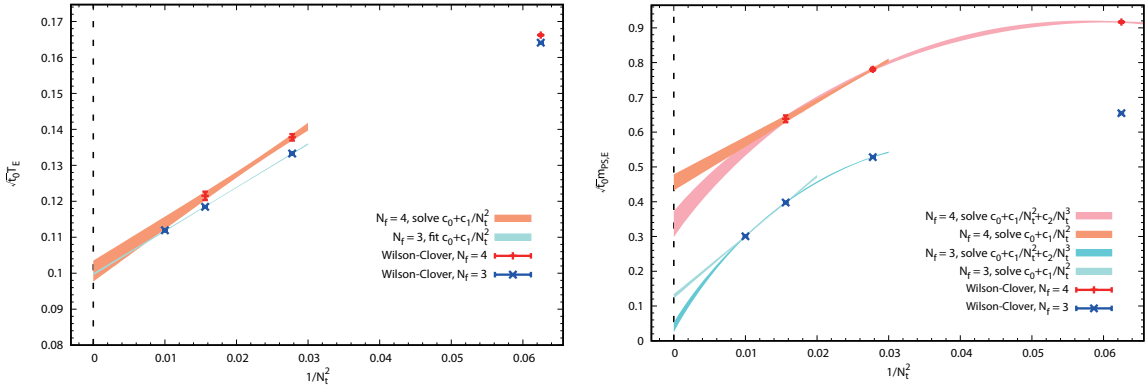}
\caption{Cutoff dependence of $\sqrt{t_0}T_E$ (left) and $\sqrt{t_0}m_\mathrm{PS,E}$ (right).
Wilson-Clover results for $N_f$ = 4 and 3 \cite{Jin:2017jjp} are shown with plus and cross symbols, respectively.
Continuum extrapolations and their statistical errors are indicated with bands.}
\label{cont_limit}
\end{center}
\end{figure}
Finally, we convert ($\beta_E$, $\kappa_E$) into physical quantities, i.e. temperature and the pseudo-scalar meson mass at the critical endpoint, ($T_E$, $m_{\mathrm{PS},E}$), where
the physical scale is determined by the $t_0$ scale given by the gradient flow method \cite{Luscher:2010iy} at zero temperature. Fig. \ref{cont_limit} summarizes cutoff dependence of
$\sqrt{t_0}T_E$ and $\sqrt{t_0}m_\mathrm{PS,E}$, where $N_f$ = 3 results with the Wilson-Clover fremions from \cite{Jin:2017jjp} are also shown.
We carry out continuum extrapolations by using polynomials of $1/N_t$ with up to a cubic term as shown in Fig \ref{cont_limit}. The left panel shows that the continuum extrapolation
of $\sqrt{t_0}T_E$ is smoothly performed and its value in the continuum limit for $N_f$ = 4 is consistent with that for $N_f$ = 3 within the statistical error. On the other hand,
it can be found from the right panel that the continuum extrapolations of $\sqrt{t_0}m_\mathrm{PS,E}$ for $N_f$ = 4 highly depend on the functional forms used, which indicates
that some of our data, especially for $N_t$ = 4, could be out of the scaling region. Therefore, simulations on finer lattices are needed for taking the continuum limit more reliably.
However, what can be clearly seen here is that $\sqrt{t_0}m_\mathrm{PS,E}$ for $N_f$ = 4 is always larger than that for $N_f$ = 3 at any $N_t$ and even in the continuum limit.

\begin{figure}[htbp]
\begin{center}
\includegraphics[width=95mm]{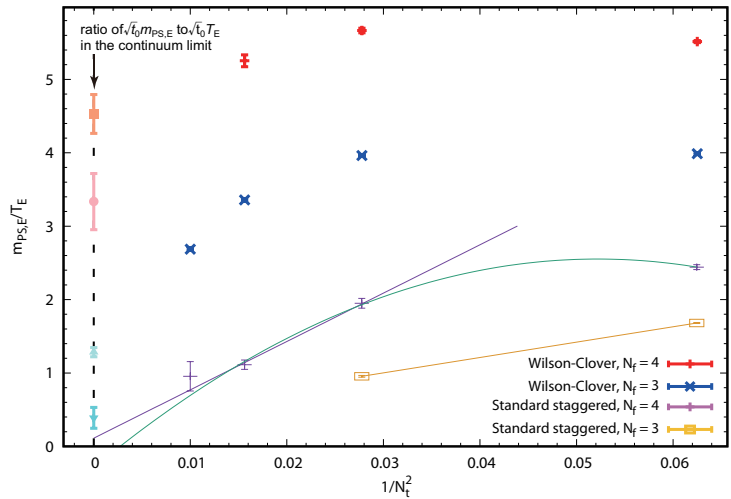}
\caption{A comparison of $m_\mathrm{PS,E}/T_E$ between Wilson-Clover and standard staggered fermions.
Results for $N_f$ = 3 with Wilson-Colver fermions and for $N_f$ = 3 and 4 with standard staggered fermions are given from \cite{Jin:2017jjp}, \cite{deForcrand:2007rq}
and \cite{deForcrand:2017cgb}, respectively, . 
The continuum extrapolated values for Wilson-Clover fermions are estimated from a ratio of $\sqrt{t_0}m_\mathrm{PS,E}$ to $\sqrt{t_0}T_E$ in the continuum limit (see texts).}
\label{cont_limit2}
\end{center}
\end{figure}
In Fig. \ref{cont_limit2} results of $m_\mathrm{PS,E}/T_E$ for $N_f$ = 3 and 4 with Wison-Clover and standard staggered fermions \cite{deForcrand:2007rq, deForcrand:2017cgb}
are compared, where the continuum extrapolated values for Wilson-Clover fermions are estimated from the ratio of $\sqrt{t_0}m_\mathrm{PS,E}$ to $\sqrt{t_0}T_E$ in the continuum limit
since the continuum extrapolations cannot be carried out smoothly due to strong cutoff dependence. With this it is found that the Wilson-Clover results are much larger than the standard
staggered ones for both $N_f$ = 3 and 4. Furthermore, the $N_f$ = 4 result seems to remain finite in the continuum limit so far in contrast to the staggered case, where $m_\mathrm{PS,E}/T_E$
is very close to zero.

\section{Conclusions}
We studied finite temperature phase transitions of $N_f$ = 4 QCD with the non-perturbatively improved Wilson-Clover fermions. We determined critical endpoints on lattices with
$N_t$ = 4, 6 and 8 by applying the kurtosis intersection method for the chiral condensate. The kurtosis value and a critical exponent for the finite-size scaling of the susceptibility peak
at the critical endpoint suggest that the second order phase transition belongs to the Z(2) universality class. We carried out continuum extrapolations of temperature and
the pseudo-scalar meson mass at the critical endpoint and found that temperature at the critical endpoint for $N_f$ = 3 and 4 are consistent with each other within statistical uncertainties.
Moreover, we showed that the pseudo-scalar meson mass at the critical endpoint for $N_f$ = 4 is larger than that for $N_f$ = 3. We also compared results with Wilson-Clover and
standard staggered fermions and found that a ratio of the pseudo-scalar meson mass to temperature at the critical endpoint for $N_f$ = 4 with Wilson-Clover fermions might remain finite
even in the continuum limit while that with standard staggered fermions is very close to zero.  Since there is no concern about the rooting for staggered ferminons for $N_f$ = 4,
the difference between Wilson and staggered fermions is at least not due to the rooting.

As we still have strong scaling violation of the pseudo-scalar meson mass at the critical endpoint in our continuum extrapolation, taking a more reliable continuum limit is needed
to obtain a concrete conclusion. Therefore, simulations on finer lattices are our future plan.

\acknowledgments{
Our numerical calculations were done on computational resources of HA-PACS, COMA and Oakforest-PACS provided by Interdisciplinary Computational
Science Program in Center for Computational Sciences at University of Tsukuba and HOKUSAI GreatWave (project ID:G17031 and G18018) at RIKEN.}

\end{document}